# Characterising molecules for fundamental physics : an accurate spectroscopic model of methyltrioxorhenium derived from new infrared and millimetre-wave measurements

Pierre Asselin,[a,b] Yann Berger [a,b]

[a]*Sorbonne Universités,UPMC Univ Paris 06,UMR 8233,MONARIS, F-75005, Paris, France*

[b]*CNRS, UMR 8233, MONARIS, F-75005, Paris, France, pierre.asselin@upmc.fr*

Thérèse R. Huet,[c] Laurent Margulès,[c] Roman Motiyenko[c]

[c]*Univ. Lille, CNRS, UMR 8523 - PhLAM - Physique des Lasers Atomes et Molécules, F-59000 Lille, France.*

Richard J. Hendricks,[d,*] Michael R. Tarbutt[d]

[d]*Centre for Cold Matter, Blackett Laboratory, Imperial College London, Prince Consort Road, London SW7 2AZ, UK*

Sean K. Tokunaga[e,f], Benoît Darquié[e,f]

[e]*Université Paris 13, Sorbonne Paris Cité, Laboratoire de Physique des Lasers, F-93430 Villetaneuse, France*

[f]*CNRS, UMR 7538, LPL, F-93430 Villetaneuse, France, benoit.darquie@univ-paris13.fr*

Precise spectroscopic analysis of polyatomic molecules enables many striking advances in physical chemistry and fundamental physics. We use several new high-resolution spectroscopic devices to improve our understanding of the rotational and rovibrational structure of methyltrioxorhenium (MTO), the achiral parent of a family of large oxorhenium compounds that are ideal candidate species for a planned measurement of parity violation in chiral molecules. Using millimetre-wave and infrared spectroscopy in a pulsed supersonic jet, a cryogenic buffer gas cell, and room temperature absorption cells, we probe the ground state and the Re=O antisymmetric and symmetric stretching excited states of both $CH_3{}^{187}ReO_3$ and $CH_3{}^{185}ReO_3$ isotopologues in the gas phase with unprecedented precision.

By extending the rotational spectra to the 150-300 GHz range, we characterize the ground state rotational and hyperfine structure up to $J$ = 43 and $K$ = 41, resulting in refinements to the rotational, quartic and hyperfine parameters, and the determination of sextic parameters and a centrifugal distortion correction to the quadrupolar hyperfine constant. We obtain rovibrational data for temperatures between 6 and 300 K in the 970-1015 $cm^{-1}$ range, at resolutions down to 8 MHz and accuracies of 30 MHz. We use these data to determine more precise excited-state rotational, Coriolis and quartic parameters, as well as the ground-state centrifugal distortion parameter $D_K$ of the $^{187}Re$ isotopologue. We also account for hyperfine structure in the rovibrational transitions and hence determine the upper state rhenium atom quadrupole coupling constant $eQq'$ .

[*] now at National Physical Laboratory, Teddington, TW11 0LW, UK.

## I- Introduction

There is an increasing demand for precision spectroscopy of polyatomic molecules. For example, precise knowledge of molecular constants is required for modeling atmospheres, whether it be our own[1,2] or those found in exo-planets[3] and in the interstellar medium[4], for studying collision physics[5,6], and as crucial groundwork for performing fundamental physics tests[7,8,9,10,11,12,13,14,15,16,17].

One such test is the measurement of the parity-violating (PV) energy difference between opposite enantiomers of a chiral molecule, induced by the weak interaction.[18] The idea that parity violation might be manifest in chiral species dates back to the 1970s.[19] A precise measurement could contribute to the determination of several parameters of the standard model of particle physics,[20] provide calibrations of relativistic quantum chemistry calculations,[18,21] and shed light on the question of biomolecular homochirality.[22]

To date, the most sensitive searches for parity violation in chiral molecules have been comparisons of transition frequencies $\nu_L$ and $\nu_R$ of the same rovibrational transition of left and right enantiomers.[23] The current upper bound on such effects is $\Delta\nu_{PV} = |\nu_L - \nu_R| < 8$ Hz, corresponding to a fractional sensitivity of $2\times10^{-13}$, obtained back in 2000 at Laboratoire de Physique des Lasers (LPL) by saturated absorption spectroscopy of the C-F stretch of CHFClBr at $\nu$~30 THz.[7,24] Subsequent theoretical studies revealed that CHFClBr's $\Delta\nu_{PV}$ was in fact very small (~2 mHz[25]), but that chiral complexes of heavy metals,[26] and in particular chiral derivatives of methyltrioxorhenium (MTO)[27], are expected to exhibit far more attainable $\Delta\nu_{PV}$'s of up to 1 Hz.

These shifts are large enough to measure, and to do so we are constructing a molecular beam Ramsey interferometer to compare the vibrational frequencies of the two enantiomers. This method has been used previously at LPL to reach resolutions of ~100 Hz and measure the absolute vibrational frequency of $SF_6$ to 0.6 Hz.[15] The PV measurement is a differential frequency measurement and we expect to reach a precision well below 0.1 Hz.[18] We have also recently demonstrated unprecedented frequency stabilities and accuracies with both $CO_2$ lasers and quantum cascade lasers[28,29,30] giving us the necessary frequency stability to make the measurement. To produce the molecular beams, we need enantiopure samples of the relevant molecules. Our collaborators have focused on chiral derivatives of MTO, a commercially available achiral complex employed in catalysis, which is a solid at room temperature and has good sublimation characteristics suitable for making a molecular beam. They have recently synthesized enantiopure MTO chiral derivatives which shows similar volatility to MTO.[27]

To take advantage of the benefits of these molecules for the PV measurement, we need to build accurate spectroscopic models of these novel species and select the best transition to use for the measurement. Precise measurements and accurate models of such large and complex heavy-metal-containing molecules are rare. To build such a model we need to (i) obtain a good understanding of all levels of structure by using a variety of spectroscopic methods that together give broadband and high resolution data across a wide spectral region, and (ii) find a procedure that transforms this large data set into a useful model. These two key advances are reported in this work.

Previously, we studied the ground and first excited states of the $\nu_{as}$ antisymmetric Re=O stretching mode of MTO. In that work[31] we reported microwave spectra obtained using a Fourier transform microwave (FTMW) spectrometer, Fourier transform infrared (FTIR) and laser absorption experiments both in cells and in supersonic beams. By combining these data, we constructed a model for MTO. In the present work, we conduct millimeter wave (MMW) and mid-infrared spectroscopy of gas phase MTO at the highest resolutions, in both room temperature cells and cold molecular beams, using a terahertz frequency multiplication chain, quantum cascade lasers and frequency stabilised $CO_2$ lasers. We also show that MTO survives laser ablation and report buffer-gas-cooling of laser-ablated gas phase MTO in a 6 K helium buffer gas cell. These new data have improved sensitivity, resolution and spectral coverage leading to definite assignments of both the antisymmetric ($\nu_{as}$) and symmetric ($\nu_s$) Re=O stretching rovibrational spectra and to a substantial refinement of the model. We use the new data to improve the accuracy of some spectroscopic constants and to determine new higher-order interaction terms. Importantly, we establish a working procedure for extracting the most precise molecular parameters from a large amount of spectroscopic data spanning centimeter to infrared wavelengths, taken at different resolutions from a variety of different experiments. This procedure can be repeated for each chiral candidate. Equally importantly, each experiment allows us to study the molecule's optical, chemical and thermodynamic properties. As we expect some of these properties to be similar for the derivatives, this is important groundwork for designing the parity violation experiment.

The paper is organized as follows. Section II describes new apparatus used for this work. These comprise the incorporation of an external-cavity quantum cascade laser (EC-QCL) into a pulsed supersonic jet setup, a distributed-feedback quantum cascade laser (DFB-QCL) and its use for acquiring spectra from laser-ablated molecules in a cryogenic cell and a new millimeter wave spectrometer. Section III describes new spectra recorded in the 150-300 GHz range, which contribute to an improved understanding of the vibrational ground state with parameters derived up to the sextic centrifugal distortion contributions. Section IV reports full analyses of new vibrational spectra of the $\nu_{as}$ antisymmetric and $\nu_s$ symmetric Re=O stretches recorded in the mid-infrared. Those were obtained *via* laser spectroscopy of molecules cooled in a pulsed jet, a skimmed continuous molecular beam and a cryogenic helium buffer gas cell, or *via* FTIR spectroscopy in a long path absorption cell at room temperature. Section IV also discusses the resulting refinements made to the model and to the analysis of the hyperfine structure (HFS) of the upper vibrational state of MTO.

## .II- Experimental details

MTO powder (from Strem Chemicals Inc., 98% purity) has been used without purification in all the experiments described in this paper.

### II-1 *Millimetre-wave experiments:* room temperature millimetre-wave spectroscopy in a cell

The spectrum of MTO was recorded in the 150-300 GHz range on the Lille millimeter-wave spectrometer[32] at room temperature (294 K), with a vapour pressure of about 20 μbar. The frequency source is a 20-GHz synthesizer (Agilent, model E8257D), stabilized on the Global Positioning System. The output frequency (12.5-18.4 GHz) is multiplied and amplified by an active sextupler AMC-10 (Virginia Diodes Inc.) providing an output power of +14 dBm in the W-band ranging from 75 to 110 GHz. Passive Schottky multipliers (×2, ×3, ×5, ×3×2, ×3×3, Virginia Diodes Inc.) are used in the last stage of the frequency multiplication chain in order to provide a useful signal in the 150-990 GHz spectral range. The absorption cell is a stainless-steel tube (6-cm diameter, 220-cm path length). In order to improve the signal sensitivity, the source is frequency modulated at 10 kHz and second harmonic detection in a lock-in amplifier is carried out. Absorption signals are detected with an InSb liquid He-cooled bolometer (QMC Instruments Ltd.). The absolute accuracy of the line-center's frequency is estimated to be better than 30 kHz for isolated lines, but worsen to 100 kHz for blended or very weak lines.

## II-2 *IR experiments*

### II-2-1 The extended cavity quantum cascade laser spectrometer coupled to a pulsed jet

The $\nu_{as}$ band of MTO has been recorded at high resolution using an extended cavity quantum cascade laser (EC-QCL) based spectrometer that includes a multipass absorption device recently developed at MONARIS. The experimental setup coupling this mid-infrared spectrometer to a pulsed supersonic jet is shown in Fig. 1. The apparatus consists in three main parts: 1) The laser and optical system, 2) The vacuum system and supersonic jet, 3) The data acquisition electronics and software.

#### *II-2-1-a The laser and the optical system*

The light source is a continuous-wave room-temperature mode-hop-free EC-QCL which covers the 930-990 $cm^{-1}$ range (Daylight Solutions). The QCL chip and a diffraction grating mounted on a piezoelectric transducer (PT) form an external cavity, and high-resolution measurements are obtained by scanning the length of this cavity. This is done by applying a sine wave with an amplitude of up to 80 V to the PT at frequencies up to 100Hz via an external controller (MDT693B, Thorlabs).

About 7.5% of the light is sent through an etalon consisting of a 0.025 $cm^{-1}$ free-spectral-range confocal Fabry–Perot cavity, to provide a relative frequency scale. A linear interpolation of the positions of the etalon maxima establishes the relationship between the voltage applied to the PT and the relative frequency. Absolute laser frequencies are obtained by sending 7.5% of the light through a reference cell containing a known reference gas. A toroidal mirror (500-mm radius of curvature) mode matches the 85 % of the remaining laser beam to a multipass near-concentric Perry cavity[33] (made of two spherical mirrors with a 125-mm radius of curvature) placed inside the vacuum chamber. The laser light crosses the supersonic expansion 20 times and focuses midway between the two mirrors at each path. Three liquid-nitrogen-cooled HgCdTe detectors (Fermionics Corporation PV 12-1 and Judson J15D12) are used to measure the powers transmitted through the multipass cavity, the etalon and the reference gas cell.

### *II-2-1-b The vacuum system and supersonic jet*

The 400 mm diameter jet chamber (Kurt J. Lesker Company) is evacuated by a combination of a 2000 l/s oil diffusion pump (Edwards, Diffstak 250/2000), a roots blower (Sogev AR350) and a rotary pump (Edwards E2M40). The diffusion pump is equipped with water-cooled baffles in order to minimize back streaming of the oil vapour towards the multipass optics. The molecular jet is produced using a pulsed circular (1-mm diameter) nozzle from General Valve Series 9 model, controlled by a valve driver (Iota One, Parker Hannifin). MTO is seeded in the supersonic jet using a brass reservoir heated to 370 K (after a design from NIST[34] and PhLAM[35]). The reservoir is filled with 0.5 g of MTO powder, and the sublimated vapour is swept away by argon at a backing pressure of 1.2 bar. Pulsed expansions are produced at a repetition rate of typically 1 Hz. The distance between the nozzle and the waist collectively formed by the reflected infrared beams inside the multipass cavity is between 3 and 7 mm.

### *II-2-1-c The data acquisition electronics and software*

Our rapid scan scheme is similar to previous designs developed for high resolution molecular spectroscopy.[36,37] The experiment is controlled by a computer running a Labview code. Spectra are recorded by driving the PT of the EC-QCL diffraction grating with a 100 Hz sine wave. This allows the QCL frequency to be scanned once per period over 0.8 $cm^{-1}$ in 5 ms. During this time window, the intensities transmitted through the supersonic jet, the etalon and the reference gas cell are digitized (12-bit acquisition card NI-PCI-6110, 5 MS/s) and recorded. A baseline-free transmittance through the multipass cavity is obtained by taking the ratio of signals recorded in presence and absence of the jet (by opening or not the nozzle).

The data acquisition sequence is illustrated in Fig. 2: the experiment starts with a TTL pulse generated from the timer card (NI-PCI-6602) that triggers both the subsequent laser sweeps (*Diffraction grating PT voltage* in Fig. 2) and 5-ms duration data acquisitions (*Acquisition triggers* and *Data acquisition* in Fig.2). It also sends a *Nozzle driver pulse* with delay $d_1$ to the Iota One controller to open the pulsed valve. For obtaining a good coincidence between the gas pulse, a laser sweep and the data acquisition, $d_1$ has to be chosen by taking into account $d_2$, the molecules' travelling times from the nozzle to the laser beams. A sequence typically consists of four background scans (in the absence of molecular jet), two taken before (scans 1 and 2 in Fig. 2) and two after (scans 4 and 5 in Fig. 2) an absorption scan (scan 3 in Fig. 2, in presence of the jet). The recorded scans are then transferred from the acquisition card to the computer to be processed and saved. This sequence is repeated at a user-defined rate of typically 1 Hz.

### *II-2-1-d Performances of the EC-QCL/pulsed jet spectrometer*

The laser rapid-scan data show a remarkable improvement in signal-to-noise ratio (by a factor between 5 and 10) over the jet FTIR spectrum of our previous work.[31]

The resolution (full-width-at-half-maximum, FWHM, of isolated rovibrational lines) of the jet-cooled QCL spectrum reported in this work (see Fig. 6 and 7 later) is about 120 MHz, limited by the Doppler broadening that results from the circular nozzle expansion geometry. † This is an improvement on the continuous jet-FTIR $\nu_{as}$ spectrum of our previous work[31], where the resolution was 160 MHz, resulting from a combination of a similar Doppler broadening (due to the use of the same geometry) and the instrumental resolution of 100 MHz. Furthermore, compared to continuous expansions, the higher backing pressures achievable with pulsed expansions allows us to obtain better rovibrational cooling. From the analysis of section IV-2-1, we estimate the MTO jet rotational temperature to be 6 ± 1 K, substantially lower than in the continuous supersonic expansion of our previous work (~10K). In our conditions, a good signal-to-noise ratio was obtained after 5000 gas pulses of 5 ms duration. This corresponds to a consumption of 200 mg of MTO, which is about 75 times less than for the continuous jet.

### II-3 Room temperature FTIR spectroscopy in a cell

We recorded both the $\nu_{as}$ and $\nu_s$ bands of the two isotopologues of $CH_3ReO_3$ (centered at about 976 and 1004 $cm^{-1}$, respectively) using a room temperature absorption cell with a path length of 150 m that has already been described elsewhere.[38] The cell was coupled to a high resolution FTIR spectrometer (IFS 125 HR, Bruker) with a maximal resolution of 0.00102 $cm^{-1}$. For those experiments a globar source, a KBr beam splitter and an HgCdTe detector were used. Because of the large difference in intensity of the two Re=O stretching bands – $\nu_{as}$ is calculated to be about 14 times more intense than $\nu_s$[39] – the antisymmetric band was recorded after injecting 0.005 mbar of MTO for an acquisition time of about 36 hours while the symmetric one was recorded by injecting ~0.014 mbar for an acquisition time of about 49 hours.

Aside from focusing on these two stretching modes, we also recorded a wide infrared vibrational spectrum from 50 to 5000 $cm^{-1}$, which will be presented elsewhere.

### II-4 The DFB QCL spectrometer coupled to a 6 K buffer-gas cell

Details of this buffer-gas cell and associated laser setup are to be found elsewhere,[40] and are thus only briefly reviewed here. A copper cell is filled with ~$10^{-2}$ mbar of helium and is cooled to 6 K using a cryocooler (Sumitomo RDK-415D). A target containing MTO is ablated with 30 mJ pulses of 8 ns from an Nd:YAG laser at 1064 nm. The resulting plume is rapidly cooled by collisions to the helium temperature, permitting spectroscopy at low temperatures for a few milliseconds, the typical diffusion time of molecules to the cold cell walls, where they freeze and are lost from the experiment. Light from a free-running Peltier-cooled (to typically -10°C) DFB QCL laser emitting at ~976 $cm^{-1}$ was sent through the cell at right angles with the plume, and absorption spectra were recorded. In order to reconstruct a frequency axis, some of the light was also sent through reference cells containing ethene or methanol, and some through a temperature-stabilized germanium reference etalon. These signals allowed for an *a posteriori* reconstruction of a frequency scale with an absolute accuracy of about 30 MHz, and a relative uncertainty of a few megahertz. [40]

---

† FWHM as narrow as 50 MHz have recently been obtained on the same set-up equipped with a pulsed slit nozzle for a $SF_6$ expansion seeded in argon.

This setup was used to investigate several regions of the $\nu_{as}$ mode of $CH_3ReO_3$. We obtain a resolution of ~8 MHz (FWHM), limited by a combination of laser frequency fluctuations, collisional and Doppler broadenings.[40] We estimate the temperature of the molecules to be 6 ± 3 K[40] (see Fig. 8 later).

### II-5 $CO_2$ laser linear absorption spectroscopy of a supersonic beam

Again, details of the setup can be found elsewhere. [9,31] A mixture of MTO and He is prepared in an oven at ~370 K. This mixture is sent to a 100 μm nozzle, through which it is expanded into vacuum to form a continuous supersonic beam. Downstream of a skimmer, a $CO_2$ laser crosses this beam at right angles for linear absorption spectroscopy. In order to enhance the signal to noise ratio, the light crosses the beam 18 times using a multi pass cell. A rotating slotted disk was used to chop the molecular beam, resulting in a ~1 kHz modulation of the absorption, detected with a lock-in amplifier. The $CO_2$ laser is locked via a phase lock loop to a second $CO_2$ laser, which is in turn locked to a saturated absorption signal of $OsO_4$ in the vicinity of the $R(20)$ $CO_2$ emission line. Using this locking scheme, we know the absolute frequency of the laser to within less than 100 Hz and obtain a stability at 1 s and a line width of respectively 1 Hz and 10 Hz.[41,42] With this apparatus, we probed a region of the $\nu_{as}$ transition of MTO at around 976 $cm^{-1}$. From the analysis in section IV-2-3, we estimate the MTO beam rotational temperature to be about 15 K (see Fig. 9 later).

## III- Millimetre-wave rotational spectroscopy: Results and analysis

Compared to our previous work, we extend here the rotational analysis of the MTO spectrum in the millimeter wave region. The rotational and hyperfine structure of both $CH_3{}^{187}ReO_3$ and $CH_3{}^{185}ReO_3$ is characterized up to $J = 43$ and $K = 41$ in the ground state (with $J$ and $K$, the quantum numbers associated with the total orbital angular momentum and its projection on the molecular symmetry axis respectively). An example spectrum is presented in Fig. 3. The ground state of both isotopologues is characterized using a standard Hamiltonian that takes into account the rotational structure ($B$), a few quartic ($D_J$, $D_{JK}$) and sextic ($H_J$, $H_{JJK}$) centrifugal distortion parameters, the spin-rotation constants ($C_{aa}$, $C_{bb}$), and the rhenium atom hyperfine quadrupole coupling constant ($eQq$), and its centrifugal distortion correction ($eQqJ$). The SPFIT/SPCAT suite of programs of H. Pickett is used to determine molecular parameters from the frequencies of assigned transitions.[43] 966 and 530 lines have been assigned for the $^{187}Re$ and $^{185}Re$ isotopologues, respectively. Of these, 46 and 19 lines for which the difference between experimental and calculated frequency is bigger than 3 times the standard deviation have been rejected in the analysis. The corresponding line lists (with the rejected lines removed) are available as ESI (ESI1). A new set of parameters has been obtained and is presented in Table 1. The value of $A$ was fixed to the value obtained in the structural determination of Wikrent et al.[44] The standard deviation of the difference between experimental and calculated frequencies is about 40 kHz, close but higher than the experimental accuracy because of the presence of several overlapped lines. Note that 21 rotational lines assigned to each isotopologue in the microwave (μW) study reported in our previous work[31] have been included in the present analysis.‡

## IV- Mid-infrared rovibrational spectroscopy: Results and analysis

### IV-1 Experimental results

We have recorded new high resolution spectra of both the symmetric $\nu_s$ and antisymmetric $\nu_{as}$ Re=O stretching bands of MTO.

Both the $\nu_s$ and $\nu_{as}$ modes have been probed at a resolution of 0.00102 $cm^{-1}$ using the room-temperature cell-FTIR setup described in Section II-3. Fig. 4 shows the full $\nu_s$ band while Fig. 5 shows a zoom of five rotational progressions of the *R*-branch. $\nu_s$ being a parallel band, it exhibits evident rotational progressions in *J* which makes it possible to directly interpret the room temperature cell data and assign transitions to it. On the other hand, a direct interpretation of the room temperature $\nu_{as}$ perpendicular band spectrum is out of reach. MTO has four vibrational modes below 600 $cm^{-1}$ and, being heavy, possesses small rotational constants. Its room temperature $\nu_{as}$ spectrum is therefore very dense given the large number of occupied rotational states and the presence of several hot bands.[45] Typical rotational progressions in *K* are thus blurred. We therefore resort to the supersonic jet technique in which the rovibrational cooling simplifies the spectrum. Fig. 6 shows the jet-cooled EC-QCL spectrum of the full $\nu_{as}$ band recorded with the setup described in Section II-2-1 and Fig. 7 is a zoom of rotational progressions of the *P*-branch.

Moving to higher resolution, linear absorption spectra of laser ablated MTO buffer-gas-cooled to ~6 K have been recorded as described in Section II-4. Fig. 8 shows an example spectrum obtained by tuning the free-running DFB QCL frequency to the *P*-branch of the $\nu_{as}$ mode, where isolated rovibrational lines can be observed. The resolution is about 8 MHz. As illustrated in Fig. 8, this allows neighbouring rovibrational lines of the two isotopologues to be resolved, an improvement on the two spectroscopic studies of supersonic jet-cooled MTO mentioned in the previous paragraph. The hyperfine structure is also partially resolved, and will be discussed below. The inset of Fig. 8 shows spectroscopy of the *Q*-branch of the $^{187}$Re MTO $\nu_{as}$ mode (here, the splitting in *J* is far smaller and is thus unresolved).

Finally, Fig. 9 shows examples of $CO_2$ laser linear absorption spectra of a continuous helium supersonic beam seeded with a few percent of MTO as described in Section II-5. The structures on the low frequency side correspond to the $^{R}Q(J,2)$ band, for *J* between 2 and 10. At these low *J* values, the expected regular rovibrational structure is strongly modified by the hyperfine interaction between the rhenium nuclear quadrupole moment and the gradient of the electric field, resulting in a dense and complex spectrum. On the high frequency side, high *J* transitions of the $^{R}Q(J,1)$ band exhibit reduced hyperfine splitting, and the amplitudes are weaker in accordance with the Boltzmann distribution at 15 K. As will be shown later, these data will be used to refine our model and estimate *eQq'*, the quadrupole coupling constant in the excited vibrational state.

‡Amongst those, only 16 lines for the $^{187}$Re isotopologue and 17 lines for the $^{185}$Re one have been kept for the present analysis (see ESI1).

**IV-2 Rovibrational analysis of the Re=O stretching bands**

MTO is a prolate symmetric top with $C_{3v}$ symmetry and 18 vibrational modes[46]. In the 1000 cm$^{-1}$ region two stretching modes are infrared active: the $\nu_{as}$(E) and $\nu_s$($A_1$) bands measured at 959 and 998 cm$^{-1}$ respectively in the solid phase[47], the rovibrational analysis of which is detailed in the following sections.

Experimental spectra of MTO are analyzed with the prolate symmetric rovibrator model, including the rotational constants $A$ and $B$, the centrifugal distortion constants $D_J$, $D_{JK}$ and $D_K$ and sextic constants $H_J$ and $H_{JJK}$ . Ground state rotational energy levels are given by:

$$E(J,K) = BJ(J+1) + (A-B)K^2 - D_J J^2(J+1)^2 - D_{JK}J(J+1)K^2 \\ -D_K K^4 - H_J J^3(J+1)^3 - H_{JJK}J^4(K+1)^2 \quad (1)$$

For the $\nu_{as}$ perpendicular band, rovibrational energy levels are calculated from the following expression:

$$E(\nu_{as},J,K) = \nu_{as} + B'J(J+1) + (A'-B')K^2 - 2\xi A'Kl \\ -D'_J J^2(J+1)^2 - D'_{JK}J(J+1)K^2 - D'_K K^4 - H'_J J^3(J+1)^3 - H'_{JJK}J^4(K+1)^2 \quad (2)$$

where $\xi$ is the first-order Coriolis constant, characteristic of a degenerate vibration. Transitions follow the selection rules $\Delta J = 0, \pm 1$; $\Delta K = \pm 1$; $\Delta l = \pm 1$ and $\Delta|K-l| = 0$ (with $l$, the quantum number associated to the vibrational angular momentum). The $\nu_{as}$ perpendicular band exhibits different kinds of transitions designated by ${}^{P,R}P(J,K)$ ${}^{P,R}Q(J,K)$ and ${}^{P,R}R(J,K)$ where the superscripts $P, R$ correspond to $\Delta K = -1, +1$, the main capital letter $P, Q, R$ to $\Delta J = -1,0,1$, respectively[48], and where $J$ and $K$ are the ground state rotational quantum numbers.

For the $\nu_s$ parallel band, rovibrational energy levels are calculated from the following expression:

$$E(\nu_s,J,K) = \nu_s + B'J(J+1) + (A'-B')K^2 - D'_J J^2(J+1)^2 - D'_{JK}J(J+1)K^2 \\ -D'_K K^4 - H'_J J^3(J+1)^3 - H'_{JJK}J^4(K+1)^2 \quad (3)$$

Transitions follow the selection rules $\Delta J = 0, \pm 1$ ; $\Delta K = 0$. Rotational lines for the $\nu_s$ parallel band are designated by ${}^{Q}P(J,K)$, ${}^{Q}Q(J,K)$ and ${}^{Q}R(J,K)$ where the superscript $Q$ means $\Delta K = 0$.

The lower-state statistical weights is affected by the $2J + 1$ rotational degeneracy factor and the rules of nuclear spin statistics for a $C_{3v}$ molecule possessing three identical H atoms of nuclear spin $I_H$ =1/2 outside the axis of symmetry. This results in transitions originating from rotational levels (in the ground vibrational state) with $K$= 3n which are twice as strong as those with $K$ =3n ± 1 (n is an integer).

As for the MMW rotational analysis of Section III, the SPFIT/SPCAT suite of programs is used[41] in the following analyses, while simulated spectra are plotted using the PGOPHER program.[49] Note also that the ground state value of $A$ is again fixed to the value obtained in the structural determination of Wikrent et al[44]

***IV-2-1 The $\nu_{as}$ spectrum***

The rovibrational structure of the jet-cooled $\nu_{as}$ spectrum (Fig. 6) which extends between 973.8 and 978.5 cm$^{-1}$ is composed of two similar rovibrational band contours centered at 976.0 and 976.6 cm$^{-1}$, respectively, with a 10.5:6 intensity ratio estimated from the integrated intensities of $Q$ branches. This corresponds to the isotopic abundance of the rhenium atom (62.93% for $^{187}$Re and 37.07% for $^{185}$Re), allowing us to assign each isotopologue a band contour. Those perpendicular bands consist of two types of rotational structures. The central part exhibits a series of strong $^{P,R}Q(J,K)$ branches for which $K$ is fixed (see also inset of Fig. 8). Each of these $K$ branches is composed of a number of unresolved lines, each corresponding to a different value of $J$, with $J \geq K$. Further from the band center, there are several progressions composed of single isolated rovibrational $^{P,R}P(J,K)$ and $^{P,R}R(J,K)$ lines.

In these $P$ and $R$ regions, the QCL-jet $\nu_{as}$ spectrum displays characteristic clusters of lines equally spaced in frequency by about 0.2 cm$^{-1}$ which are likely to correspond to $^{P}P(J,K)$ or $^{R}R(J,K)$ transitions for which $J$ is fixed. We start by assigning the most intense lines of these clusters to the stronger $^{P}P(J,3n)$ and $^{R}R(J,3n)$ transitions. We note that the lines' frequency is an increasing function of the $K$ quantum number in $^{P}P(J,K)$ clusters, while it is a decreasing function in $^{R}R(J,K)$ clusters. Equations (1) and (2) show that this happens when the $K$-dependent term $(-2\xi A')$ is much larger than the $K^2$-dependent term $(A'-A)-(B'-B)$. The near zero value of the Coriolis coupling constant $\xi$ of our previous work[31] (which had been determined from the simulation of the $Q$-branch alone) contradicts our observations and is now refuted. Theoretical predictions using $\xi$ = 0.20 derived from a DFT (B3LYP) calculation nicely confirm our observations.

All intense patterns of the jet-cooled spectrum are then assigned to the $^{187}$Re isotopologue. Most of unassigned lines are likely to correspond to the less abundant $^{185}$Re isotopologue. As shown in Fig. 7, an almost perfect overlap is observed between the $^{P}P(J,K)$ transitions of the $^{185}$Re isotopologue and the $^{P,R}P(J-3,K-3)$ lines of the $^{187}$Re one.

The rovibrational lines assigned up to $J$ = 10 at that stage provide a first set of new constants including one ground-state parameter ($D_K$, which was not accessible from the µW/MMW experiments) and 4 excited-state parameters ($\nu_{as}$, $A'$, $B'$ and $\xi$) for each isotopologue. In these preliminary analyses, ground-state rotational and both ground- and excited-state centrifugal distortion parameters are fixed to the µW/MMW values of Table 1.

We now return to the room temperature spectrum, for which the model is now good enough to assign higher $J$ lines. 240 (37) lines for $CH_3{}^{187}ReO_3$ ($CH_3{}^{185}ReO_3$) have been identified, reaching quantum numbers of up to $J$=33 (17) for $^{P}P$ transitions and 25 (12) for $^{R}R$ transitions. Note that 15 lines for $CH_3{}^{187}ReO_3$ (5 for $CH_3{}^{185}ReO_3$) were observed in the $R$-branch of both the jet and cell/FTIR spectra. In the final line list (see ESI2), for these lines, the more precise frequencies extracted from the cell/FTIR data (30 MHz resolution) have been preferred.

Having assigned these higher $J$ lines, we now focus on $P$-branch transitions where $^{185}$Re and $^{187}$Re isotopologues transitions are nearly overlapped in both supersonic jet and cell/FTIR spectra (see Fig. 7). Although the absolute accuracy using the buffer-gas setup is comparable to the other methods, the relative uncertainty of the frequency scale is only a few megahertz. These lines are thus resolved in the buffer-gas-cooled spectra, (see Fig. 8). The positions of 6 jet-cooled $^{P}P$ lines of $^{187}$Re MTO, and 7 jet-cooled $^{P}P$ and $^{R}P$ lines of $^{185}$Re MTO have thus been refined and replaced by frequencies extracted from buffer-gas-cooled spectra (those are labeled in the final line lists given in ESI2). We point out that 5 amongst the 7 $CH_3{}^{185}ReO_3$ lines are now fully resolved and assigned a position different (shifted by about 100 MHz) from the neighbouring $^{187}$Re isotopologue line frequency.

Finally, a global fit to a total of 323 line positions was performed for the $^{187}$Re isotopologue. The assignment of higher $J$ lines allow the extraction of yet 3 additional parameters, namely excited-state $D_J'$, $D_{JK}'$, $D_K'$ centrifugal distortion parameters. The set of parameters obtained at that stage is given in set 1 of Table 2.

A similar fit to a total of 96 lines was realized for $CH_3{}^{185}ReO_3$, and led to set 3 in Table 2. Here centrifugal distortion parameters of both the ground and excited state were constrained to the μW/MMW ground-state values. Only $D_K'$ could be floated while $D_K$ was fixed to the $^{187}$Re MTO ground-state value of the previous analysis. Note that including the buffer-gas lines allows us to better reproduce the relative line positions between isotopologues (see Fig. 8).

Fig. 7 shows the good agreement between a simulated spectrum of the $\nu_{as}$ mode of both isotopologues with a rotational temperature of 6 ± 1 K and jet-cooled experimental data for rotational progressions in the 974.8-975.6 cm$^{-1}$ range. Fig. 6 also shows our best agreement between the data and a simulated stick spectrum over the entire band at the same temperature.

### *IV-2-2 The $\nu_s$ spectrum*

The $\nu_s$ band is observed in the cell-FTIR spectrum recorded at the maximal resolution of 0.00102 cm$^{-1}$ (see Sections II-3 and IV-1). The whole transition spans 40 cm$^{-1}$ from 985 to 1025 cm$^{-1}$ but only the 995-1015 cm$^{-1}$ range shown in Fig. 4 has been considered in the analysis. The $^{Q}P(J,K)$ branch is strongly perturbed by red shifted hot bands and the nearby $\nu_{as}$ band centered at 976 cm$^{-1}$ but the characteristic $PQR$ structure of the $\nu_s$ parallel band of a symmetric top molecule is still visible with resolved $K$ substructures in the $P$- and $R$-branches. 848 (296) lines have been assigned to the $^{187}$Re ($^{185}$Re) isotopologue, with highest $J$ values equal to 51 (30) for $^{Q}R$ transitions and 40 (30) for $^{Q}P$ transitions. Five rotational parameters ($\nu_s$, $A'$, $B'$, $D_J'$ and $D_{JK}'$) in the $v_s = 1$ state have been adjusted to reproduce the assigned transition frequencies of both isotopologues.

As for the $\nu_{as}$ band of $CH_3{}^{185}ReO_3$, only $D_K'$ was adjusted independently while $D_K$ was fixed to the $^{187}$Re MTO ground-state value of the previous analysis. The results of those two least squares optimizations are reported in set 1 and set 3 of Table 3, for the $^{187}$Re and $^{185}$Re isotopologues respectively. The final line lists for the $\nu_s$ band is reported in ESI3. Fig. 5 shows the good agreement between the data and a simulation of the $\nu_s$ modes of both isotopologues for five rotational progressions of the $R$-branch with resolved $K$ lines. Fig. 4 also shows good agreement between the data and a simulated stick spectrum on a broader span.

### *IV-2-3 Including the hyperfine structure*

The HFS of MTO has not been taken into account in the above analyses of the Re=O vibrational modes. Microwave[44,50] and millimeter-wave experiments show that the rhenium quadrupole coupling constant $eQq$ > 700 MHz is relatively close in magnitude to the rotational constants, resulting in the break-up of microwave and millimeter-wave transitions into hyperfine components. The HFS must be included in the analysis of the most resolved data.

In particular, to refine our understanding of the HFS in the excited vibrational state of the $\nu_{as}$ perpendicular band of $CH_3{}^{187}ReO_3$, we proceed to the assignment of the structures observed in the continuous supersonic beam $CO_2$ laser absorption spectra displayed in Fig. 9. As mentioned above, these structures are strongly modified by the hyperfine interaction. In a parallel work[40], an analytical hyperfine Hamiltonian which included the rhenium electric quadrupole interaction and spin-rotation magnetic interactions was fitted to four ${}^PP$ lines from the buffer-gas setup (including the ${}^PP(6,3)$ line in Fig. 8). The quadrupole coupling constant in the upper state, $eQq'$, was found to be 716(3) MHz, to be compared to $eQq$ = 716.5725(26) in the ground state (see Table 1), and so we conclude that there is little variation of the quadrupole coupling constant with vibration. To assign hyperfine components to the structures observed in the supersonic beam linear absorption spectra of Fig. 9, we thus compare the data to a simulation that uses both the rovibrational parameters of set 1 in Table 2 as well as the HFS parameters obtained from μW/MMW spectroscopy (Table 1), using the latter for both the ground and excited vibrational states. From this comparison, a total of 36 hyperfine components of the ${}^RQ(J,2)$ lines with $2 \leq J \leq 10$ and of the ${}^RQ(16,1)$ line have been assigned, with frequency uncertainties ranging from $1\times10^{-4}$ cm$^{-1}$ to $3\times10^{-4}$ cm$^{-1}$, depending on the width of the assigned structure. These have been added to the previous $^{187}$Re MTO $\nu_{as}$ band line list (see ESI2, where they are labeled).

For both the $\nu_s$ and $\nu_{as}$ bands of $^{187}$Re and $^{185}$Re isotopologues of MTO, we now fit and simulate the transition frequencies with a Hamiltonian containing the rhenium quadrupole coupling constants $eQq$ and its centrifugal distortion correction $eQqJ$, and the spin-rotation interaction terms $C_{aa}$ and $C_{bb}$ in both the ground and excited states. For the $^{187}$Re (and $^{185}$Re) isotopologue, we have to fit line lists including: (i) the 16(17) assigned rotational lines from the microwave study reported in our previous work[31] (ii), the 920(511) assigned rotational lines at higher values of $J$ in the MMW study of section III, and (iii), the most intense hyperfine components (those with $\Delta F = \Delta J$) of all the rovibrational transitions assigned in section IV-2. Note that hyperfine components of a given rovibrational line are unresolved and thus are all assigned the same observed frequency. For the particular case of the $^{187}$Re MTO $\nu_{as}$ mode, the 36 partially resolved hyperfine components assigned in the spectrum of Fig. 9 are added to the line list. A total of 2891 (respectively 6024) ${}^PP$, ${}^RP$ and ${}^RR$ lines are assigned in $J, K$ and $F$ to the $\nu_{as}$ (respectively $\nu_s$) band of the $^{187}$Re isotopologue, and 1101 (respectively 2304) lines are assigned to the $\nu_{as}$ (respectively $\nu_s$) band of the $^{185}$Re isotopologue. The complete lists of assigned lines are available as ESI for the two modes of each isotopologue.

For each line list, we perform a fit in which we float all parameters but $A$ ($D_K$ is also fixed in 3 amongst the 4 cases). However, with the exception of the upper-state quadrupole coupling constant of the $\nu_{as}$ mode of $^{187}$Re MTO, all hyperfine parameters and sextic centrifugal distortion constants were constrained to the same values for both ground and excited states. Note that, for the $\nu_{as}$ band of $CH_3{}^{185}ReO_3$ $D'_J$ and $D'_{JK}$ were also constrained to be equal in both states due to the limited number of lines. The four corresponding sets of molecular parameters are gathered in sets 2 and 4 of Table 2 and 3. For the $\nu_{as}$ mode of $^{187}$Re MTO, it was also possible to float $eQq'$, after the assignment of hyperfine components to the structures in Fig. 9. We obtained $eQq'$ = 705(44) MHz, which is in good agreement with our analysis of the buffer-gas-cooled spectra detailed in Ref. 40, confirming the very weak variation of the quadrupole coupling constant with vibration.

Fig. 7, 8 and 9 compare data and simulations that include HFS. Although this makes little difference for the EC-QCL spectra in Fig 7 where the resolution is similar to the HFS splitting (~100MHz), it is indispensable for obtaining agreement for the $CO_2$ laser (Fig. 9) and buffer gas spectra (Fig. 8), at a rotational temperature of 6 K and 15 K, respectively.

## V- Conclusion

In this paper, we report an in-depth spectroscopic study of MTO, in the context of molecular PV measurement. The most promising route for this measurement is to perform Ramsey interferometry on a molecular beam, and the chiral derivatives of MTO are excellent candidate species, because calculations predict PV frequency shifts large enough to measure, and because it is possible to synthesize these molecules and bring them into the gas phase.

A full characterization of such organometallic molecules requires the use of several sensitive spectroscopic techniques, much like the experiments reported here. Three new MMW and IR high resolution spectroscopic devices coupled either to a room temperature cell, a cryogenic cell or a pulsed supersonic expansion, were used to fully reinvestigate the spectrum of MTO in the ground state and Re=O stretching $\nu_s$ and $\nu_{as}$ excited states. The large dataset obtained was then used to refine the rovibrational model of MTO. A full set of rotational and hyperfine parameters of the $\nu_s$ and $\nu_{as}$ modes of both isotopologues $^{187}$Re and $^{185}$Re of MTO, including centrifugal distortion contributions, has been extracted from the combined analysis of different μW, MMW and IR spectra. The assignment of more than thirty partially resolved hyperfine transitions in $CO_2$ laser absorption spectra of a skimmed molecular beam led to a determination of the quadrupole coupling constant in the excited state.

In order to select the best transition in which to measure $\Delta\nu_{PV}$, a similar effort will be required for the chiral derivatives. Nonetheless, we establish a procedure for extracting very accurate molecular parameters of complex species, both in the ground and excited state, from a large amount of spectroscopic data spanning various spectral regions, taken at different resolutions from a variety of experiments. Studying MTO also provides insight into the physical and chemical properties of these large and complex novel species, which is important for designing the molecular beam Ramsey interferometer.

Finally, beyond our scope of interest, a similar sort of analysis is becoming increasingly necessary for any precise measurement on polyatomic molecules, whether it be for atmospheric, planetary or interstellar studies, chemistry, biology or other fundamental physics tests.

## Acknowledgments

The authors are grateful to the AILES staff and particularly to Olivier Pirali for providing access to the Bruker high resolution IFS 125 spectrometer on the AILES beamline, and for the optical settings of the long path White-type cell coupled to it. The authors also thank A. Amy-Klein, C. Chardonnet, C. Daussy and E.A. Hinds for fruitful discussions. In France, this work was supported by ANR (under grants no ANR 2010 BLAN 724 3, no ANR-12-ASTR-0028-03 and no ANR-15-CE30-0005-01, and through Labex First-TF ANR 10 LABX 48 01), Région Île-de-France (DIM Nano-K), CNRS, Université Paris 13 and AS GRAM. In the UK, the work was supported by EPSRC under grant EP/I012044/1. The work was made possible through the International Exchanges Programme run jointly by the Royal Society and CNRS.

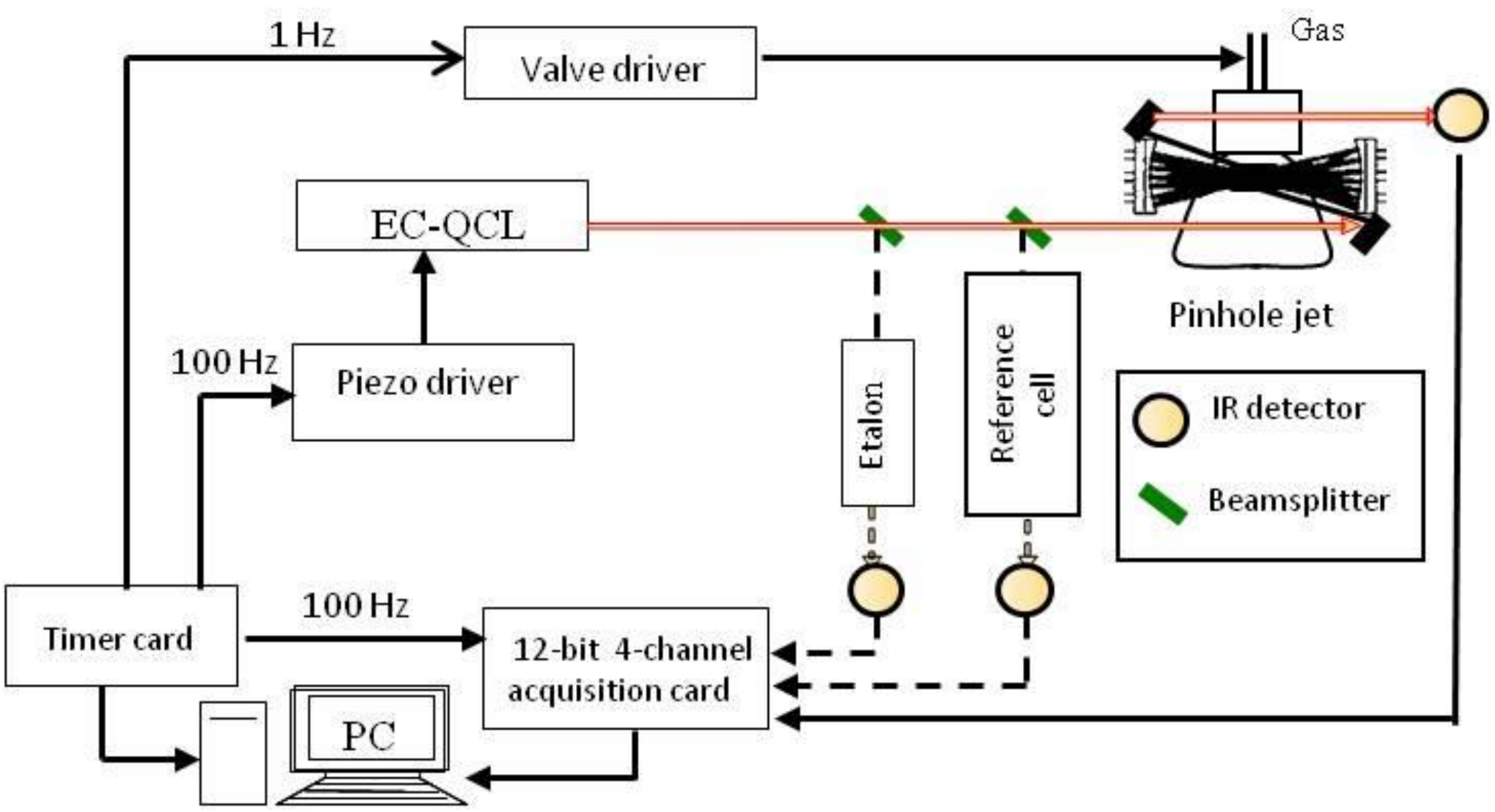

Experiment control

Figure 1: Experimental setup combining a pulsed supersonic jet originating from a circular nozzle and an EC-QCL based spectrometer that includes a near concentric multipass optics. EC-QCL: external-cavity quantum cascade laser.

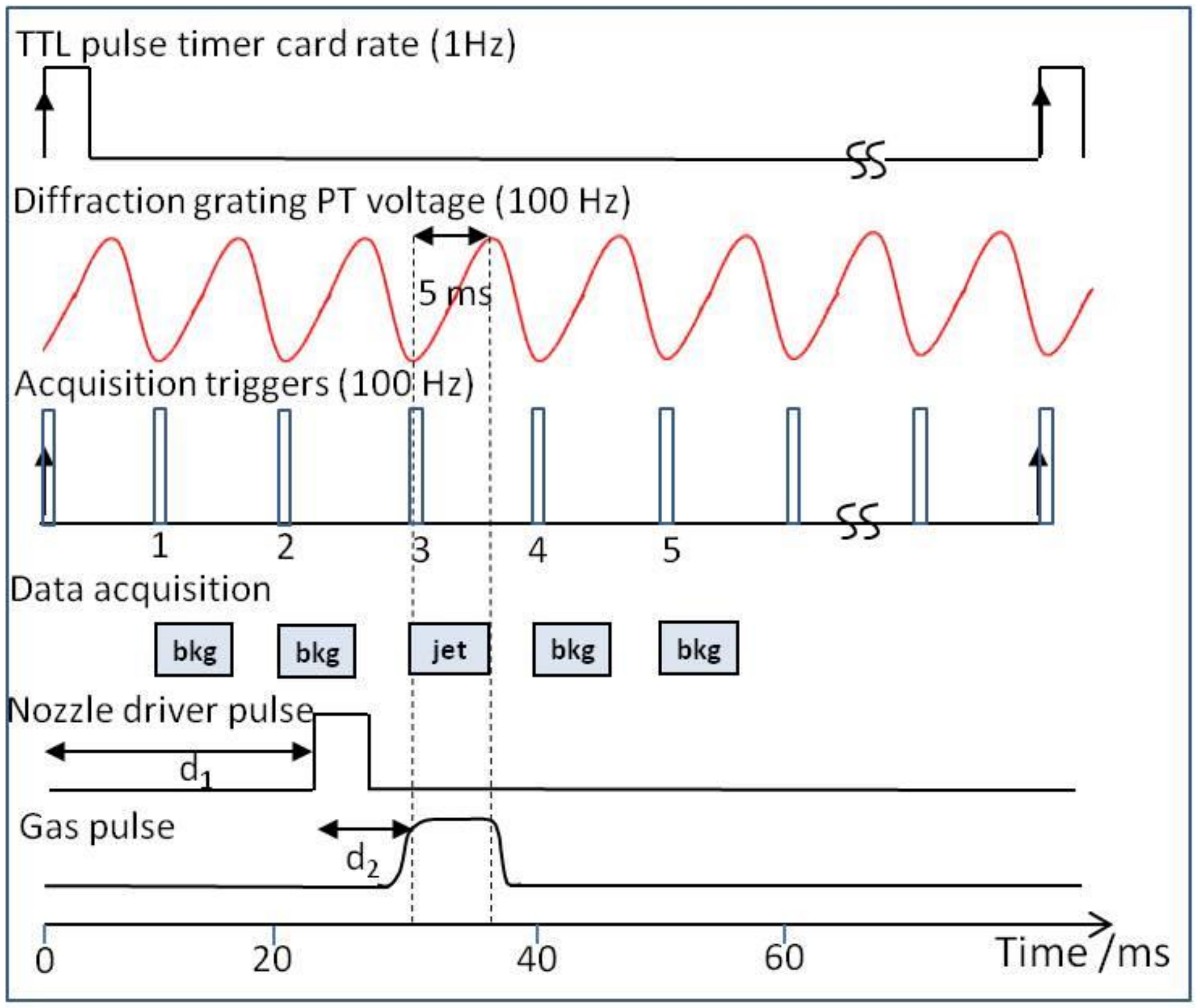

Pulse sequence

Figure 2: Data acquisition sequence for the jet-cooled EC-QCL experiment. The TTL pulse timer card triggers three channels: 1) the *Diffraction grating PT voltage* sine wave , 2) the *Acquisition triggers* sequence which defines the starting time of each 5-ms acquisition period, 3) the *Nozzle driver pulse* after an adjustable delay $d_1$. After each TTL pulse the signal of only 5 (amongst 100 at a 1-Hz repetition rate) acquisition windows are kept: 4 bkg (background signal in the absence of molecular jet) and 1 jet (in the presence of the molecular jet). The delay $d_2$ is the molecules' travelling times from the nozzle to the laser beams. PT: piezoelectric transducer.

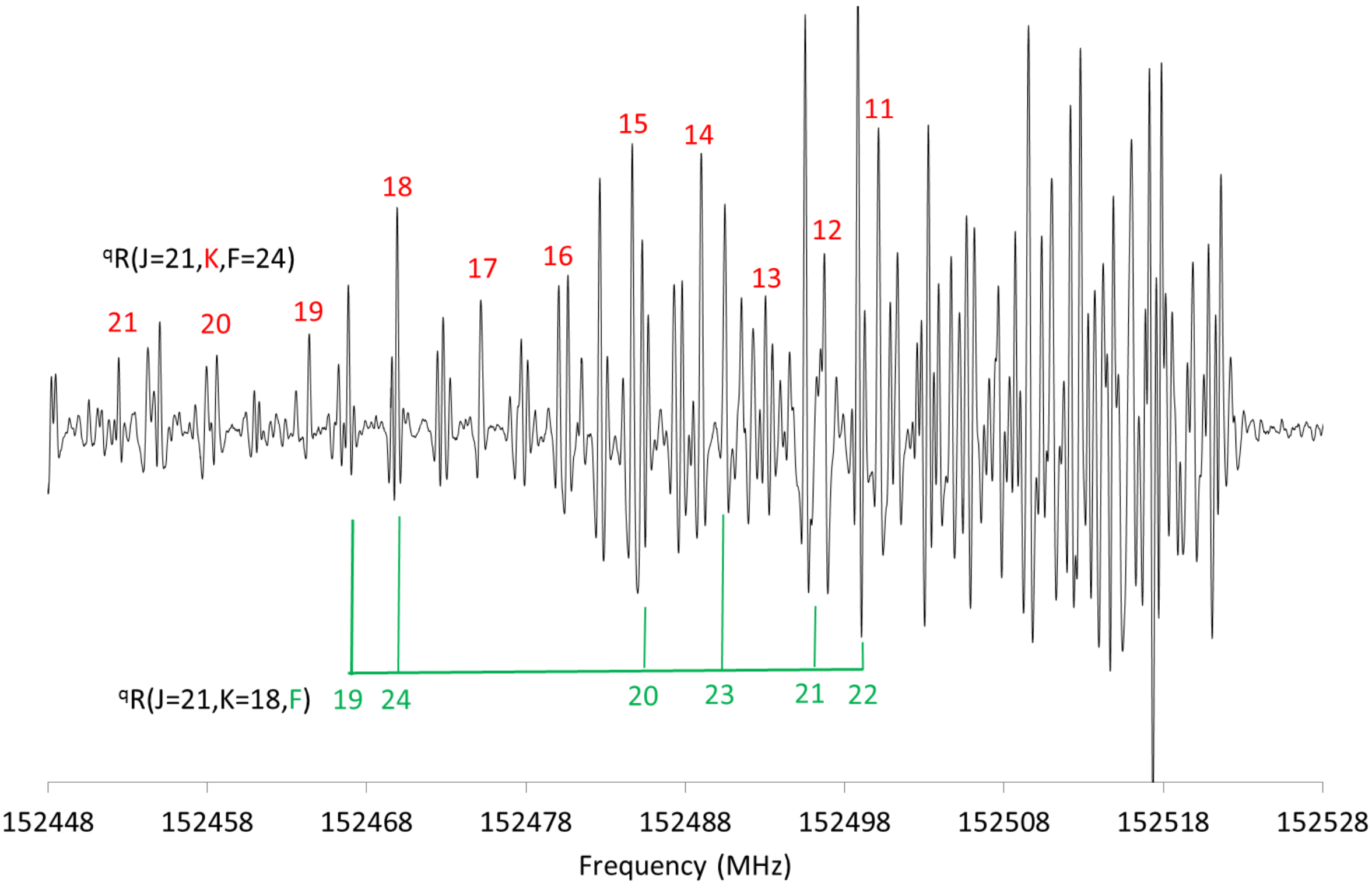

Figure 3: Small portion of the millimeter-wave spectrum illustrating the hyperfine structure observed on the rotational spectrum of $CH_3{}^{187}ReO_3$. The assignments of the $\Delta F$=+1 hyperfine components for the ${}^{Q}R(J=21,K=18,F)$ line are presented in green, as well as the $K$ structure of the $\Delta F$=+1 ${}^{Q}R(J=21,K,F=24)$ lines in red. $J$, $K$ and $F$ are the ground state rotational quantum numbers.

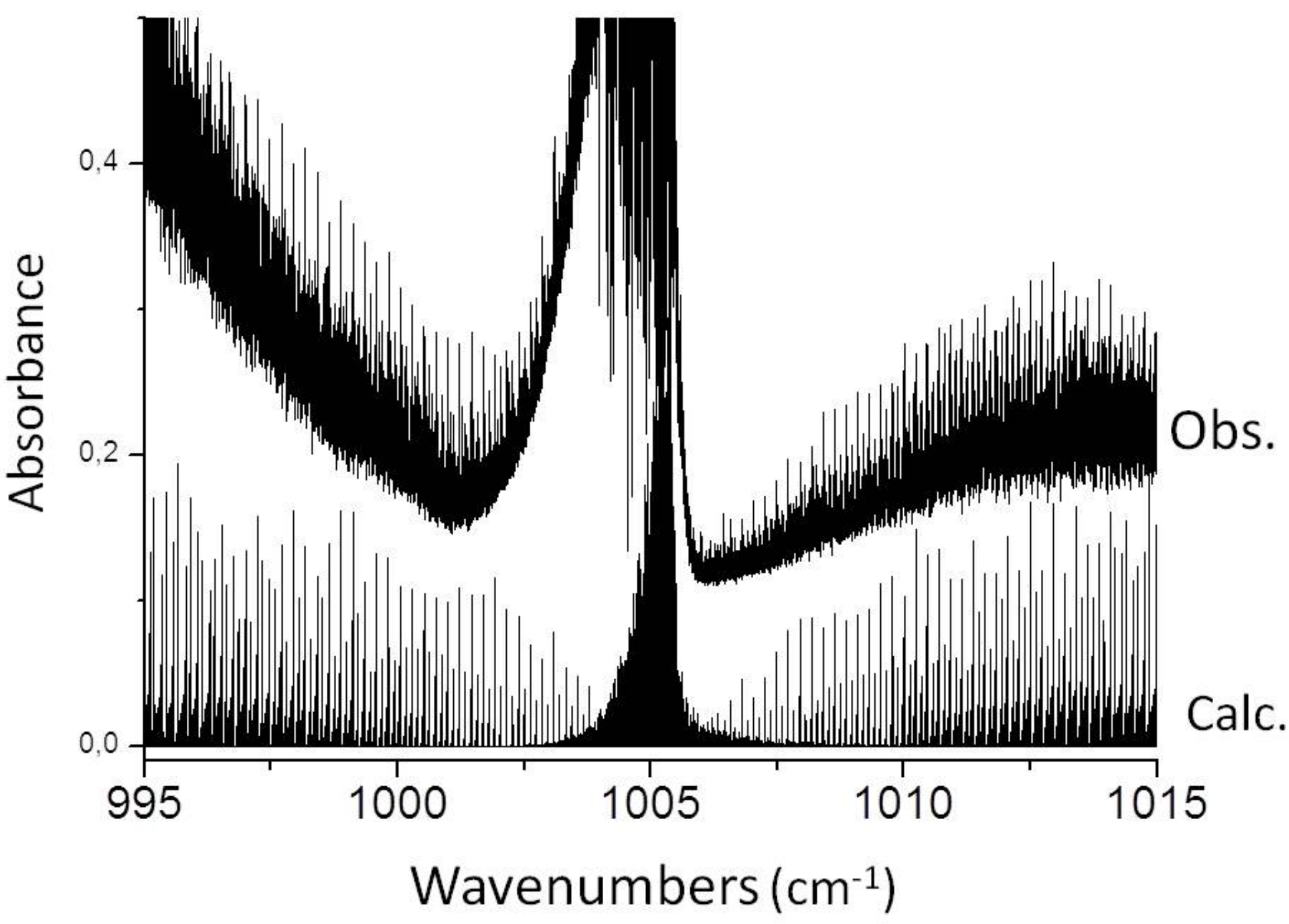

Figure 4: Comparison between the observed room temperature cell-FTIR spectrum of the $\nu_s$ band of MTO and the calculated stick spectrum (using sets of parameters 1 and 3 of Table 3, for the $^{187}$Re and $^{185}$Re isotopologues respectively). The experimental $^{Q}Q$ branch is strongly broadened by the many hot bands red shifted with respect to the band center. Note also that below 995 cm$^{-1}$, high *J* lines of $\nu_s$ *P*-branch overlap with the $\nu_{as}$ *R*-branch.

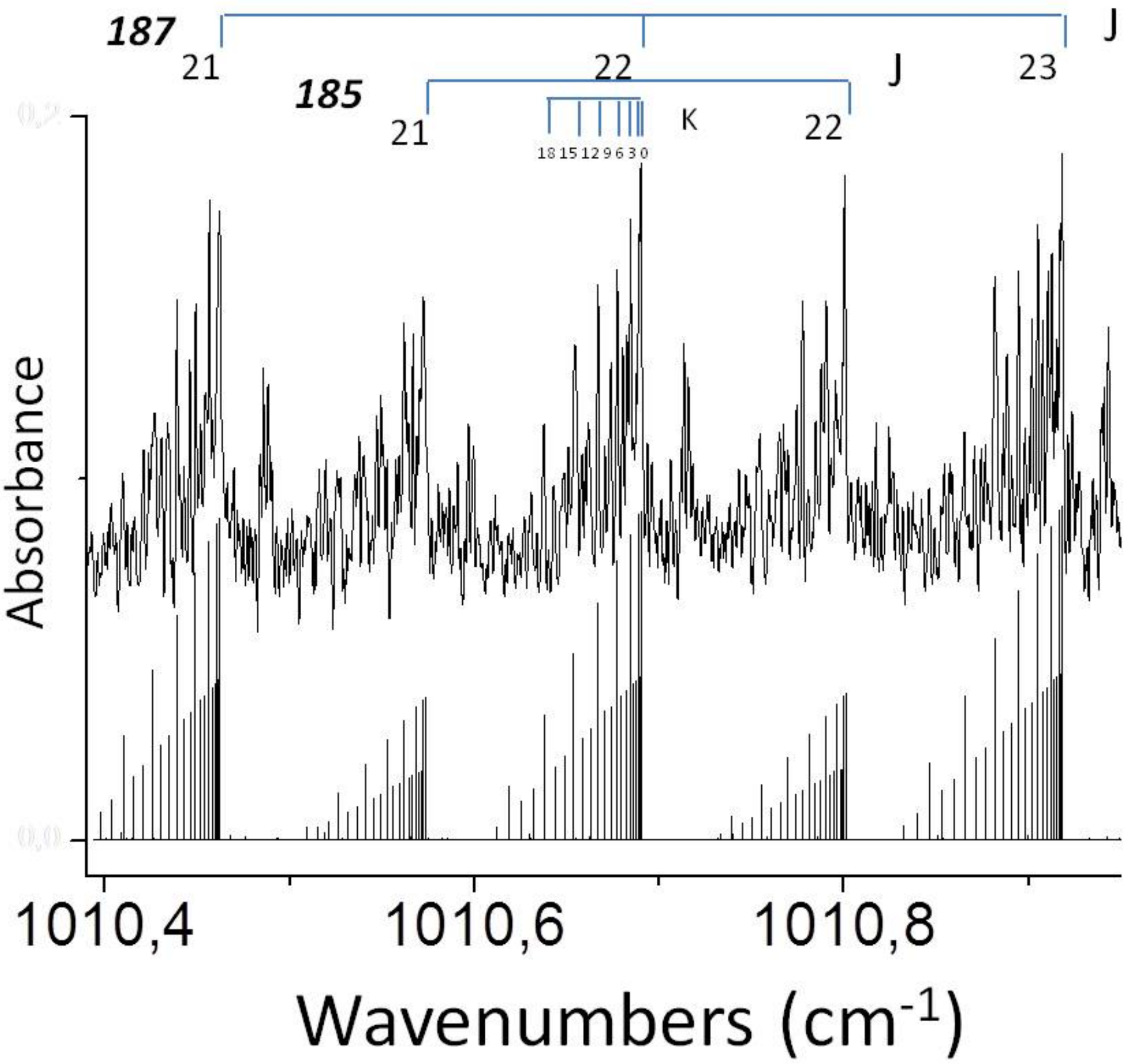

Figure 5: Comparison between observed room temperature cell-FTIR spectrum and our best calculated spectrum (using sets of parameters 1 and 3 of Table 3), with assigned rotational progressions of both isotopologues, located in the *R*-branch of the $\nu_s$ band. Those *K* progressions are composed of ${}^{Q}R(J,K)$ lines for which $J$ is fixed, each corresponding to different *K* values with $K \leq J$. Three progressions of the $^{187}$Re isotopologue for $J = 21$-23 and two progressions of the $^{185}$Re isotopologue for $J = 21$ and 22 are visible. The features observed between the assigned progressions of the fundamental of $\nu_s$ are attributed to hot bands.

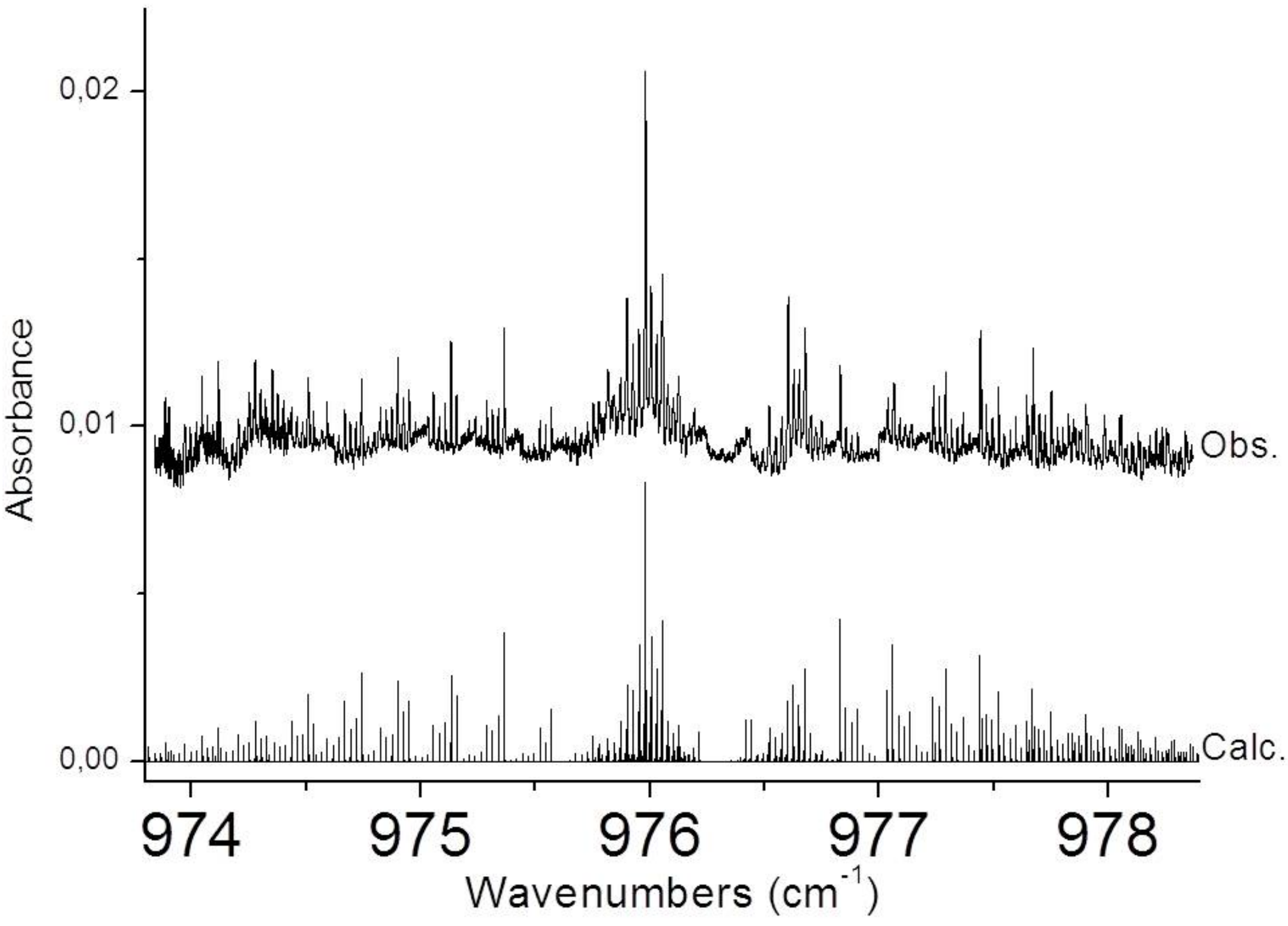

Figure 6: Comparison between the observed jet-cooled EC-QCL spectrum of the $\nu_{as}$ band and a simulated stick spectrum of the $\nu_{as}$ modes of both isotopologues without the contribution of the hyperfine structure (using set of parameters 1 and 3 of Table 2), at a rotational temperature of 6 K.

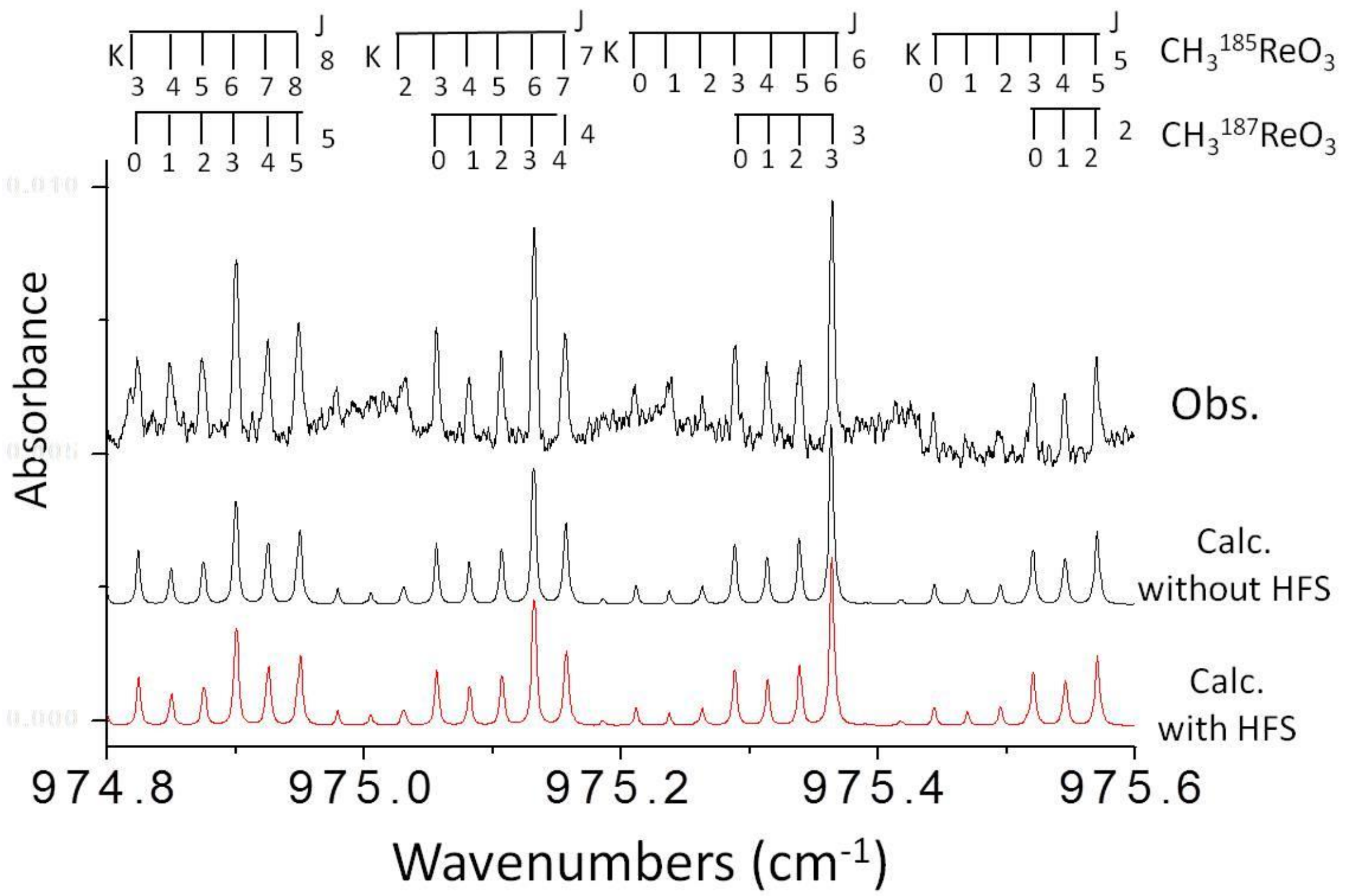

Figure 7: Comparison between the observed jet-cooled EC-QCL spectrum and two calculated spectra at 6 K, without the HFS (using set of parameters 1 and 3 of Table 2) and with the HFS (using set of parameters 2 and 4 of Table 2), for four assigned rotational progressions of both isotopologues, located in the *P*-branch of the $\nu_{as}$ band. Those *K* progressions are composed of ${}^{P}P(J,K)$ and ${}^{R}P(J,K)$ lines for which $J$ is fixed, each corresponding to a different value of *K*, with $K \leq J$. The four progressions of the ${}^{187}$Re and ${}^{185}$Re MTO are those for which $J$ = 2-5 and $J$ = 5-8 respectively. Calculated spectra are obtained by convolving stick spectra with a Lorentzian profile of 100 MHz full-width-at-half-maximum, corresponding to the expected resolution. Including the HFS does not significantly modify the simulations because the HFS splitting is of the same order as the resolution.

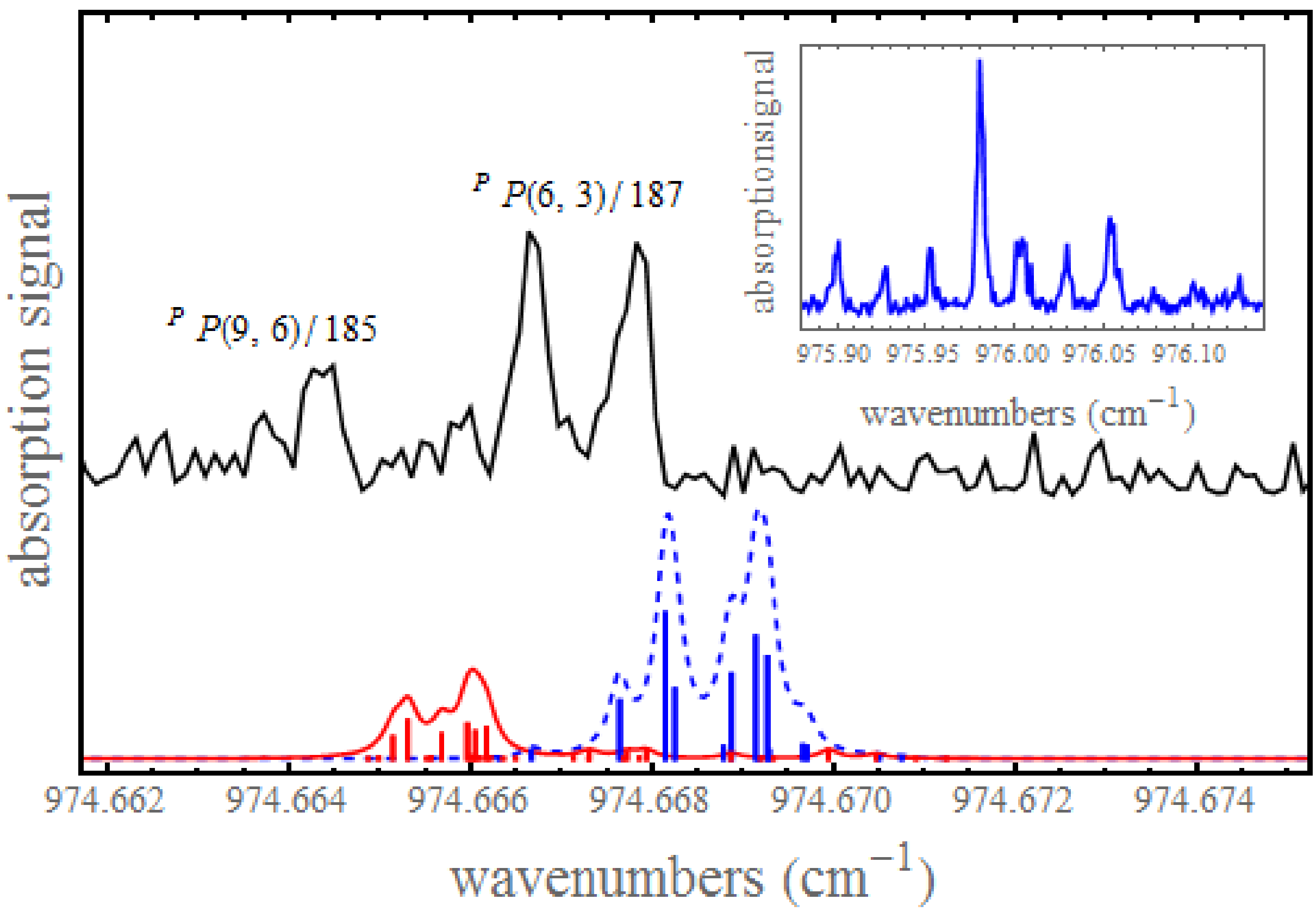

Figure 8: Upper black curve: Linear absorption spectrum of the ablation plume of MTO buffer-gas-cooled to ~6 K recorded using a free-running DFB QCL. It exhibits two isolated rovibrational lines, the $^{P}P(9,6)$ and $^{P}P(6,3)$ lines of the $^{185}$Re and $^{187}$Re MTO isotopologues respectively, both showing a partially resolved hyperfine structure. Lower dashed blue curve: simulated $CH_3{}^{187}ReO_3$ spectrum at a rotational temperature of 6 K, using set 2 of Table 2. Lower solid red curve: simulated $CH_3{}^{185}ReO_3$ spectrum at a rotational temperature of 6 K, using set 4 of Table 2. Stick spectra (blue for $CH_3{}^{187}ReO_3$ and red for $CH_3{}^{185}ReO_3$) are convolved by a Lorentzian profile of 8 MHz full-width-at-half-maximum, corresponding to the expected resolution.[40] The small offset between the experimental and simulated spectra is consistent with the 30 MHz absolute accuracy of the frequency scale[40] (see section II-4). Insert: Buffer-gas-cooled spectrum showing the rotational contour of the $Q$ branch of the antisymmetric Re=O stretching mode of the $^{187}$Re isotopologue at ~976 cm$^{-1}$ from which we infer a rotational temperature $T_{rot} = 6 \pm 3$ K.[40]

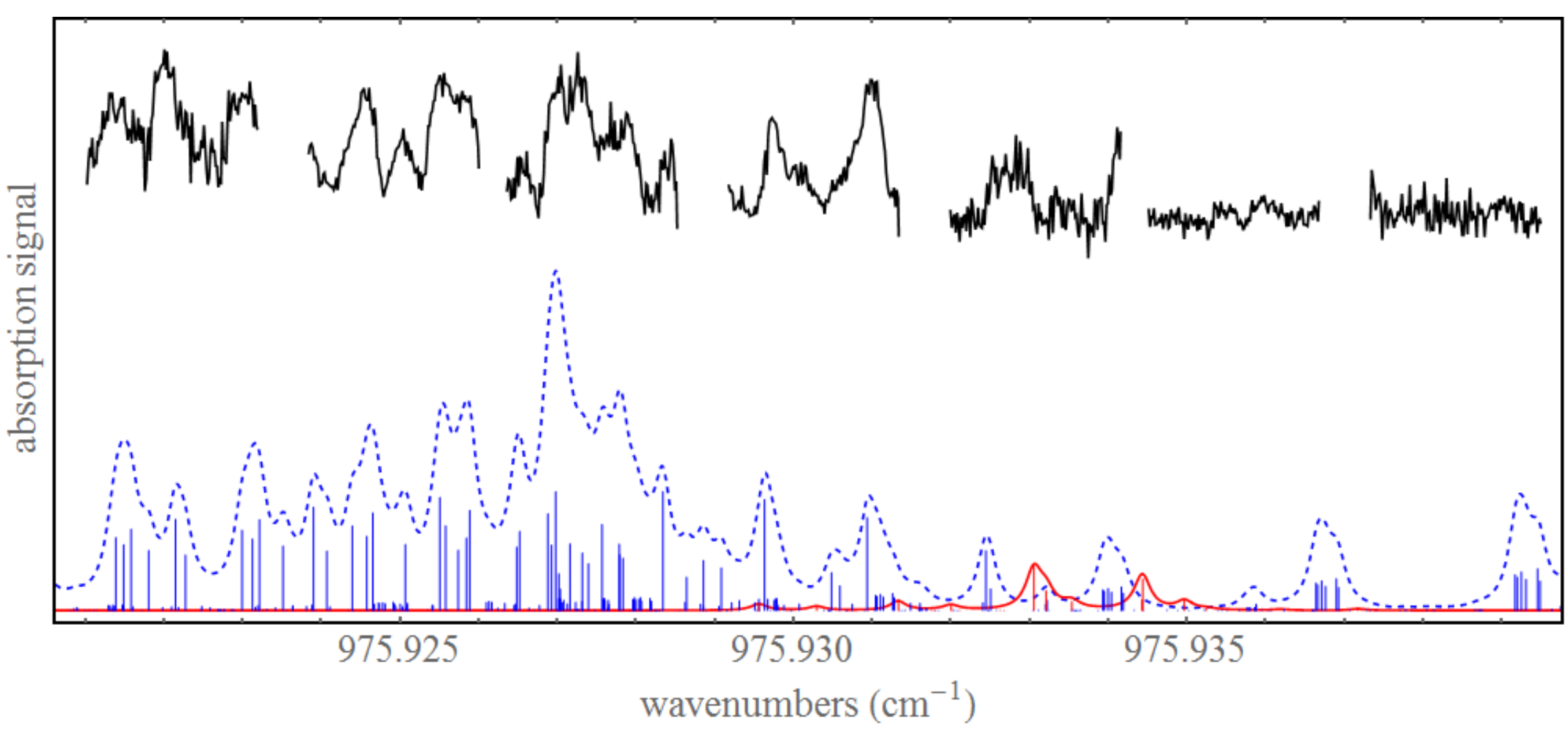

Figure 9: Upper black curve: linear absorption spectroscopy of an MTO-seeded continuous skimmed supersonic beam, in the vicinity of the *R*(20) $CO_2$ laser line frequency (located at ~975.93044 cm$^{-1}$); experimental conditions9: 100 μm nozzle diameter, 750 μm skimmer diameter, helium backing pressure ∼ 1 bar, oven temperature ∼ 90°C, translational longitudinal temperature ∼ 1 K, 9 mm nozzle-to-skimmer distance, 1 point recorded every 100 kHz, ∼ 1 s of integration time per point, $1.5\times10^{-5}$ cm$^{-1}$ binning, ∼ 0.5 μW of laser power in each of the 18 passes on the multi-pass cell (see text), ∼ 5 mm beam waist. Dashed blue curve: simulated $CH_3{}^{187}ReO_3$ spectrum at a rotational temperature of 15 K, using set of parameters 2 of Table 2. Solid red curve: simulated $CH_3{}^{185}ReO_3$ spectrum at a rotational temperature of 15 K, using set of parameters 4 of Table 2. Stick spectra (blue for $CH_3{}^{187}ReO_3$ and red for $CH_3{}^{185}ReO_3$) are convolved by a Lorentzian profile of full-width-at-half-maximum $2.5\times10^{-4}$ cm$^{-1}$, corresponding to the expected Doppler broadening resulting from the divergence of the beam.[9]

| | $CH_3{}^{187}ReO_3$ | $CH_3{}^{185}ReO_3$ |
|---|---|---|
| $A$ | 3849.81[a] | 3849.81[a] |
| $B$ | 3466.9645880(822) | 3467.0491870(705) |
| $D_J \times 10^3$ | 0.6610990(988) | 0.661262(111) |
| $D_{JK} \times 10^3$ | 2.354181(322) | 2.352230(456) |
| $H_J \times 10^9$ | 0.3204(342) | 0.2930(407) |
| $H_{JJK} \times 10^9$ | 2.8190(999) | 2.740(138) |
| $eQq$ | 716.57253(259) | 752.21433(251) |
| $eQqJ$ | -0.008453(266) | -0.008327(344) |
| $C_{aa}$ | - 0.050600(455) | - 0.050680(367) |
| $C_{bb}$ | - 0.051750(109) | - 0.051221(105) |
| Number of distinct lines / $J_{max}$ / $K_{max}$ | 608 / 43 / 41 | 313 / 43 / 38 |
| St. dev. | 0.042 | 0.037 |
| [a]Fixed value [44]. | | |

Table 1: Molecular parameters (in MHz) for the ground states of $CH_3{}^{187}ReO_3$ and $CH_3{}^{185}ReO_3$ obtained from µW and MMW data. Uncertainties corresponding to one standard deviation are indicated into parentheses. St. dev.: standard deviation (in MHz) of the difference between experimental and calculated frequencies.

| | $\nu_{as}$ | | | |
|---|---|---|---|---|
| | $CH_3{}^{187}ReO_3$ | | $CH_3{}^{185}ReO_3$ | |
| | Set 1 (without HFS) | Set 2 (with HFS) | Set 3 (without HFS) | Set 4 (with HFS) |
| $A$/MHz | 3849.81[a] | 3849.81[a] | 3849.81[a] | 3849.81[a] |
| $B$/MHz | 3466.964588[b] | 3466.964588 (74) | 3467.049187[b] | 3467.049187(71) |
| $D_J$/kHz | 0.661099[b] | 0. 661099 (89) | 0.661262[b] | 0.66126(11) |
| $D_{JK}$/kHz | 2.354181[b] | 2.354181 (290) | 2.35223[b] | 2.35223(46) |
| $D_K$/kHz | -11.5(3) | -11.748 (263) | -11.5[d] | -11.5[d] |
| $H_J$/mHz | 0.3204[b] | 0.3204 (308)[c] | 0.293[b] | 0.293(41)[c] |
| $H_{JJK}$/mHz | 2.819[b] | 2.819 (90)[c] | 2.740[b] | 2.74(14)[c] |
| $eQq$/MHz | - | 716.5725 (24) | - | 757.215(3)[c] |
| $eQqJ$/MHz | - | - 0.00845 (24)[c] | - | -0.0084(3)[c] |
| $C_{aa}$/kHz | - | - 50.60 (41)[c] | - | -50.69(37)[c] |
| $C_{bb}$/kHz | - | - 51.750 (99)[c] | - | -51.25(11)[c] |
| $\nu_{as}$/cm$^{-1}$ | 976.02062(19) | 976.0210936 (738) | 976.64044(31) | 976.64045(31) |
| $A'$/MHz | 3844.38(4) | 3844.3019 (290) | 3844.64(23) | 3844.65(23) |
| $B'$/MHz | 3464.50(5) | 3464.4671 (248) | 3464.46(7) | 3464.46(7) |
| $\xi$ | 0.1989(1) | 0.198954 (45) | 0.1995(2) | 0.1995(1) |
| $D'_J$/kHz | 0.771(77) | 0.753 (62) | 0.66126[b] | 0.66126[c] |
| $D'_{JK}$/kHz | 2.14(17) | 2.107 (164) | 2.35223[b] | 2.35223[c] |
| $D'_K$/kHz | -10.9(3) | -10.940 (300) | -11.6(10) | -11.6(10) |
| $H'_J$/mHz | 0.3204[b] | 0. 3204 (308)[c] | 0.293[b] | 0.293(41)[c] |
| $H'_{JJK}$/mHz | 2.819[b] | 2.819 (90)[c] | 2.740[b] | 2.74(14)[c] |
| $eQq'$/MHz | - | 705 (44) | - | 757.215(3)[c] |
| $eQqJ$/MHz | - | - 0.00845 (24)[c] | - | -0.0084(3)[c] |
| $C'_{aa}$ | - | - 50.60 (41)[c] | - | -50.69(37)[c] |
| $C'_{bb}$ | - | - 51.750 (99)[c] | - | -51.25(11)[c] |

[a] Fixed value [44]. [b] Fixed to the μW/MMW ground state parameters of Table 1.
[c] Fitted and constrained to the same values for both ground and excited states.
[d] Value fixed to that obtained in set 1.

Table 2: Ground- and excited-state parameters of the $\nu_{as}$ band for both MTO isotopologues. All microwave, millimeter-wave and mid-infrared experiments described in this work were used. The uncertainties in parentheses correspond to one standard deviation.

| | $\nu_s$ | | | |
|---|---|---|---|---|
| | $CH_3{}^{187}ReO_3$ | | $CH_3{}^{185}ReO_3$ | |
| | Set 1 (without HFS) | Set 2 (with HFS) | Set 3 (without HFS) | Set 4 (with HFS) |
| $A$/MHz | 3849.81[a] | 3849.81[a] | 3849.81[a] | 3849.81[a] |
| $B$/MHz | 3466.964588[b] | 3466.964589(74) | 3467.049187[b] | 3467.049187(71) |
| $D_J$/kHz | 0.661099[b] | 0.661099(89) | 0.661262[b] | 0.66126(11) |
| $D_{JK}$/kHz | 2.354181[b] | 2.35418(29) | 2.35223[b] | 2.35223(46) |
| $D_K$/kHz | -11.5[d] | -11.5[d] | -11.5[d] | -11.5[d] |
| $H_J$/mHz | 0.3204[b] | 0.320(31)[c] | 0.293[b] | 0.293(41)[c] |
| $H_{JJK}$/mHz | 2.819[b] | 2.819(90)[c] | 2.740[b] | 2.741(139)[c] |
| $eQq$/MHz | - | 716.573(2)[c] | - | 757.214(3)[c] |
| $eQqJ$/MHz | - | - 0.0085(2)[c] | - | -0.00833)[c] |
| $C_{aa}$/kHz | - | - 50.60(41)[c] | - | -50.68(37)[c] |
| $C_{bb}$/kHz | - | - 51.75(10)[c] | - | -51.22(11)[c] |
| $\nu_s$/cm$^{-1}$ | 1005.41078(11) | 1005.41078(11) | 1005.52047(22) | 1005.52043(22) |
| $A'$/MHz | 3842.89(2) | 3842.89(2) | 3842.98(11) | 3842.98(11) |
| $B'$/MHz | 3464.84(1) | 3465.031(6) | 3464.98(3) | 3465.19(4) |
| $D'_J$/kHz | 0.6575(24) | 0.6574(24) | 0.696(39) | 0.73(4) |
| $D'_{JK}$/kHz | 2.356(11) | 2.356(13) | 2.49(19) | 2.27(18) |
| $D'_K$/kHz | -11.53(3) | -11.52(4) | -11.9(5) | -11.4(5) |
| $H'_J$/mHz | 0.3204[b] | 0.320(31)[c] | 0.293[b] | 0.293(41)[c] |
| $H'_{JJK}$/mHz | 2.819[b] | 2.819(90)[c] | 2.740[b] | 2.741(139)[c] |
| $eQq'$/MHz | - | 716.573(2)[c] | - | 757.214(3)[c] |
| $eQqJ$/MHz | - | - 0.0085(2)[c] | - | -0.0083(3)[c] |
| $C'_{aa}$ | - | - 50.60(41)[c] | - | -50.68(37)[c] |
| $C'_{bb}$ | - | - 51.75(10)[c] | - | -51.22(11)[c] |

[a] Fixed value [44]. [b] Fixed to the μW/MMW ground state parameters of Table 1.
[c] Fitted and constrained to the same values for both ground and excited states.
[d] Value fixed to that obtained in set 1 of Table 2.

Table 3: Ground- and excited-state parameters of the $\nu_s$ band for both MTO isotopologues. Microwave and millimeter-wave (section III), and mid-infrared data from the room-temperature cell-FTIR experiments (section IV-2-2) were used. The uncertainties in parentheses correspond to one standard deviation.

## References

[1] M. Guinet, D. Mondelain, C. Janssen and C. Camy-Peyret, *J. Quant. Spectrosc. Radiat. Transf.* **111**, 961 (2010).

[2] J. J. Harrison, P. F. Bernath and G. Kirchengast, *J. Quant. Spectrosc. Radiat. Transf.* **112**, 2347 (2011).

[3] J. Tennyson, S. N. Yurchenko, A. F. Al-Refaie, E. J. Barton, K. L. Chubb, P. A. Coles, S. Diamantopoulou, M. N. Gorman, C. Hill, A. Z. Lam, L. Lodi, L. K. McKemmish,Y. Na, A. Owens, O. L. Polyansky, T. Rivlin, C. Sousa-Silva, D. S. Underwood, A. Yachmenev and E. Zak, *J. Mol. Spectr.* **327**, 73 (2016).

[4] E. Herbst and E. F. van Dishoeck, *Annu. Rev. Astron. Astrophys.* **47**, 427 (2009).

[5] J.-M. Hartmann, C. Boulet and D. Robert, *Collisional effects on molecular spectra: laboratory experiments and models, consequences for applications* (Elsevier Science, Amsterdam, 2008).

[6] J. Buldyreva, N. Lavrentieva and V. Starikov, *Collisional line broadening and shifting of atmospheric gases* (Imperial College Press, London, 2011).

[7] C. Daussy, T. Marrel, A. Amy-Klein, C. T. Nguyen, C. Bordé and C. Chardonnet, *Phys. Rev. Lett.* **83**, 1554 (1999).

[8] J. J. Hudson, D. M. Kara, I. J. Smallman, B. E. Sauer, M. R. Tarbutt and E. A. Hinds, *Nature* **473**, 493 (2011).

[9] S. K. Tokunaga, C. Stoeffler, F. Auguste, A. Shelkovnikov, C. Daussy, A. Amy-Klein, C. Chardonnet, and B. Darquié, *Mol. Phys.* **111**, 2363 (2013).

[10] J. Baron, W. C. Campbell, D. DeMille, J. M. Doyle, G. Gabrielse, Y. V. Gurevich, P. W. Hess, N. R. Hutzler, E. Kirilov, I. Kozyryev, B. R. O'Leary, C. D. Panda, M. F. Parsons, E. S. Petrik, B. Spaun, A. C. Vutha and A. D.West, *Science* **343**, 269 (2014).

[11] S. B. Cahn, J. Ammon, E. Kirilov, Y. V. Gurevich, D. Murphree, R. Paolino, D. A. Rahmlow, M. G. Kozlov and D. DeMille, *Phys. Rev. Lett.* **112**, 163002 (2014).

[12] S. Mejri, P. L. T. Sow, O. Kozlova, C. Ayari, S. K. Tokunaga, C. Chardonnet, S. Briaudeau, B. Darquié, F. Rohart and C. Daussy, *Metrologia* **52**, S314 (2015).

[13] J. Biesheuvel, J. Karr, L. Hilico, K. S. E. Eikema, W. Ubachs and J. C. J. Koelemeij, *Nat. Commun.* **7**, 1 (2016).

[14] E. R. Hudson, H. J. Lewandowski, B. C. Sawyer and J. Ye, *Phys. Rev. Lett.* **96**, 143004 (2006).

[15] A. Shelkovnikov, R. J. Butcher, C. Chardonnet and A. Amy-Klein, *Phys. Rev. Lett.* **100**, 150801 (2008).

[16] S. Truppe, R. J. Hendricks, S. K. Tokunaga, H. J. Lewandowski, M. G. Kozlov, C. Henkel, E. A. Hinds and M. R. Tarbutt, *Nat. Commun.* **4**, 2600 (2013).

[17] P. Jansen, H. L. Bethlem and W. Ubachs, *J. Chem. Phys.* **140**, 10901 (2014).

[18] B. Darquié, C. Stoeffler, A. Shelkovnikov, C. Daussy, A. Amy-Klein, C. Chardonnet, S. Zrig, L. Guy, J. Crassous, P. Soulard, P. Asselin, T. R. Huet, P. Schwerdtfeger, R. Bast and T. Saue, *Chirality* **22**, 870 (2010).

[19] D. W. Rein, *J. Mol. Evol.* **4**, 15 (1974).

[20] M. A. Bouchiat and C. C. Bouchiat, *Rep. Prog. Phys.* **60**, 1391 (1997).

[21] M Quack, J Stohner, and M Willeke, *Annu. Rev. Phys. Chem.* **59**, 741 (2008).

[22] C. Viedma and P. Cintas. *Isr. J. Chem.*, **51**, 997 (2011).

[23] V. S. Letokhov, *Phys. Lett. A,* **53**, 275 (1975).

[24] M. Ziskind, T. Marrel, C. Daussy and C. Chardonnet, *EPJD*, **20**, 219 (2002).

[25] P. Schwerdtfeger, T. Saue, J. N. P. van Stralen and L. Visscher, *Phys. Rev. A*, **71**, 012103 (2005).
[26] P. Schwerdtfeger and R. Bast, *J. Am. Chem. Soc.* **126**, 1652 (2004).
[27] N. Saleh, S. Zrig, T. Roisnel, L. Guy, R. Bast, T. Saue, B. Darquié and J. Crassous, *Phys. Chem. Chem. Phys.* **15**, 10952 (2013).
[28] P. L. T. Sow, S. Mejri, S. K. Tokunaga, O. Lopez, A. Goncharov, B. Argence, C. Chardonnet, A. Amy-Klein, C. Daussy and B. Darquié, *App. Phys. Lett.* **104**, 264101 (2014).
[29] B. Chanteau, O. Lopez, W. Zhang, D. Nicolodi, B. Argence, F. Auguste, M. Abgrall, C. Chardonnet, G. Santarelli, B. Darquié, Y. Le Coq and A. Amy-Klein, *New J. Phys.* **15**, 073003 (2013).
[30] B. Argence, B. Chanteau, O. Lopez, D. Nicolodi, M. Abgrall, C. Chardonnet, C. Daussy, B. Darquié, Y. Le Coq and A. Amy-Klein, *Nature Photon* **9**, 456 (2015).
[31] C. Stoeffler, B. Darquié, A. Shelkovnikov, C. Daussy, A. Amy-Klein, C. Chardonnet, L. Guy, J. Crassous, T. R. Huet, P. Soulard and P. Asselin, *Phys. Chem. Chem. Phys.* **13**, 854 (2011).
[32] O. Zakharenko, R. A. Motiyenko, L. Margulès and T. R. Huet. *J. Mol. Spectrosc*., **317**, 41 (2015)
[33] D. Kaur, A. M. Souza, J. Wanna, S. A. Hammad, L. Mercorelli and D.S. Perry, Appl. Opt. **29**, 119 (1990).
[34] R.D. Suenram, F.J. Lovas, D.F. Plusquellic, A. Lessari, Y.Kawashima, J.O. Jensen, A.C. Samuels, J. Mol. Spectrosc. 211 110 (2002).
[35] S. Kassi, D. Petitprez and G. Wlodarczak, J. Mol. Spectr. **228**, 293 (2004).
[36] M. D. Brookes, C. Xia, J. A. Anstey, B. G. Fulsom, K-X. Au Yong, J. M. King and A. R. W. McKellar, *Spectrochim. Acta Part A* **60**, 3235 (2004).
[37] X. Liu, Y. Xu, Z. Su, W. S. Tam and I. Leonov, *Appl. Phys. B*, **102**, 629 (2011).
[38] O. Pirali, V. Boudon, J. Oomens and M. Vervloet, *J. Chem. Phys.***136**, 024310 (2012).
[39] R. Bast and T. Saue, private communication.
[40] S. K. Tokunaga, R. J. Hendricks, M. R. Tarbutt, and B. Darquié, arXiv:1607.08741 [physics.atom-ph] (2016).
[41] V. Bernard, C. Daussy, G. Nogues, L. Constantin, P. E. Durand, A. Amy-Klein, A. van Lerberghe, and C. Chardonnet, IEEE J. Quantum Electron. 33, 1282 (1997).
[42] C. Chardonnet, "*Spectroscopie de saturation de hautes précision et sensibilité en champ laser fort. Applications aux molécules $OsO_4$, $SF_6$ et $CO_2$ et à la métrologie des fréquences*," Ph.D. thesis (Université Paris 13, Villetaneuse, 1989).
[43] H. M. Pickett, *J. Mol. Spectrosc.* **148**, 371 (1991).
[44] P. Wikrent, B. J. Drouin, S. G. Kukolich, J. C. Lilly, M. T. Ashby, W. A. Herrmann and W. Scherer, *J. Chem. Phys.* **107**, 2187 (1997).
[45] O. Pirali, M. Goubet, T. Huet, R. Georges, P. Soulard, P. Asselin, J. Courbe, P. Roy and M. Vervloet , *Phys. Chem. Chem. Phys.* **15**, 10141 (2013).
[46] S. F. Parker and H. Herman, *Spectrochim. Acta Part A* **56**, 1123 (2000).
[47] J. Mink, G. Keresztury, A. Stirling and W. A. Hermann, *Spectrochim. Acta Part A* **50**, 2039 (1994).
[48] G. Herzberg, *Molecular Spectra and Molecular Structure II, Infrared and Raman Spectra*, D. Van Nostrand Company Inc., New York, 1945.
[49] PGOPHER, *a program for Simulating Rotational Structure*, C. M. Western, University of Bristol, http://pgopher.chm.bris.ac.uk
[50] S. M. Sickafoose, P. Wikrent, B.J. Drouin and S. G. Kukolich, *Chem. Phys. Lett.* **263**, 191 (1996).